\RequirePackage{fix-cm}

\documentclass[twocolumn,epjc3]{svjour3}  

\usepackage{graphicx}
\usepackage[switch]{lineno}
\RequirePackage[numbers,sort&compress]{natbib}
\RequirePackage[colorlinks,citecolor=blue,urlcolor=blue,linkcolor=blue]{hyperref}
\smartqed  

\journalname{Eur. Phys. J. C}

\begin{document}

 \newcommand{\gerda}{\textsc{Gerda}}
 \newcommand{\mjd}{\textsc{Majorana Demonstrator}}
 \newcommand{\legend}{\textsc{Legend}}
 \newcommand{\bb}{$0\nu\beta\beta$}
 \newcommand{\qbb}{$Q_{\beta\beta}$}
 \newcommand{\K}{$^{42}$K }
  \newcommand{\Ba}{$^{133}$Ba }
 \newcommand{\Co}{$^{60}$Co}
 \newcommand{\Cs}{$^{137}$Cs}
 \newcommand{\Am}{$^{241}$Am } 
 \newcommand{\Amc}{$^{241}$Am, } 
 \newcommand{\Th}{$^{228}$Th } 
 \newcommand{\Thc}{$^{228}$Th, } 
 \newcommand{\Thp}{$^{228}$Th. } 
 \newcommand{\Bi}{$^{212}$Bi } 
  \newcommand{\Po}{$^{210}$Po}  
 \newcommand{\Tl}{$^{208}$Tl } 
 \newcommand{\Bii}{$^{214}$Bi}  
 \newcommand{\Tlp}{$^{208}$Tl. } 
 \newcommand{\Ge}{$^{76}$Ge }
 \newcommand{\Gep}{$^{76}$Ge. }  
 \newcommand{\Se}{$^{82}$Se } 
 \newcommand{\Mo}{$^{100}$Mo } 
 \newcommand{\Te}{$^{130}$Te } 
 \newcommand{\Xe}{$^{136}$Xe }
 \newcommand{\expo}{8.5~kg$\cdot$yr}
 \newcommand{\bidx}{4.9^{+7.3}_{-3.4}}
  
 \title{Characterization of inverted coaxial \Ge detectors in GERDA for future double-$\beta$ decay experiments}
 
 
\author{
	The \mbox{\protect{\sc{Gerda}}} collaboration\thanksref{corrauthor}
	\and  \\[4mm]
	M.~Agostini\thanksref{UCL,TUM} \and
	G.~Araujo\thanksref{UZH} \and
	A.M.~Bakalyarov\thanksref{KU} \and
	M.~Balata\thanksref{ALNGS} \and
	I.~Barabanov\thanksref{INRM} \and
	L.~Baudis\thanksref{UZH} \and
	C.~Bauer\thanksref{HD} \and
	E.~Bellotti\thanksref{MIBF,MIBINFN} \and
	S.~Belogurov\thanksref{ITEP,INRM,alsoMEPHI} \and
	A.~Bettini\thanksref{PDUNI,PDINFN}
% ...
}
\authorrunning{the \textsc{Gerda} collaboration} 
\thankstext{corrauthor}{\emph{correspondence} \texttt{gerda-eb@mpi-hd.mpg.de}}
\thankstext{alsoMEPHI}{\emph{also at:} NRNU MEPhI, Moscow, Russia}
\thankstext{alsoLev}{\emph{also at:} Moscow Inst. of Physics and Technology,
	Russia}
\thankstext{nowIKZ}{\emph{present address:} Leibniz-Institut f{\"u}r
	Kristallz{\"u}chtung, Berlin, Germany}
\thankstext{alsoDubna}{\emph{also at:} Dubna State University, Dubna, Russia} 
\thankstext{nowMacolino}{\emph{present address:} LAL, CNRS/IN2P3,
	Universit{\'e} Paris-Saclay, Orsay, France} 
\institute{
	INFN Laboratori Nazionali del Gran Sasso and Gran Sasso Science Institute, Assergi, Italy\label{ALNGS}
% ...
\and
	Max-Planck-Institut f{\"u}r Kernphysik, Heidelberg, Germany\label{HD} \and
	Department of Physics and Astronomy, University College London, London, UK\label{UCL} \and
	Dipartimento di Fisica, Universit{\`a} Milano Bicocca, Milan, Italy\label{MIBF} \and
	INFN Milano Bicocca, Milan, Italy\label{MIBINFN}
% ...
\and
	Institute for Nuclear Research of the Russian Academy of Sciences, Moscow, Russia\label{INRM} \and
	Institute for Theoretical and Experimental Physics, NRC ``Kurchatov Institute'', Moscow, Russia\label{ITEP} \and
	National Research Centre ``Kurchatov Institute'', Moscow, Russia\label{KU}
% ...
\and
	Physik Department, Technische  Universit{\"a}t M{\"u}nchen, Germany\label{TUM} \and
	Dipartimento di Fisica e Astronomia, Universit{\`a} degli Studi di 
	Padova, Padua, Italy\label{PDUNI} \and
	INFN  Padova, Padua, Italy\label{PDINFN}
% ...
\and
	Physik-Institut, Universit{\"a}t Z{\"u}rich, Z{u}rich, Switzerland\label{UZH}
}

\date{Received: date / Accepted: date}

\maketitle

\begin{abstract}
	
	Neutrinoless double-$\beta$ decay of \Ge is searched for with germanium detectors where source and detector of the decay are identical.
	For the success of future experiments it is important to increase the mass of the detectors.
	We report here on the characterization and testing of five prototype detectors manufactured in inverted coaxial (IC) geometry from material enriched to 88\% in \Gep
	IC detectors combine the large mass of the traditional semi-coaxial Ge detectors with the superior resolution and pulse shape discrimination power of point contact detectors which exhibited so far much lower mass.
	Their performance has been found to be satisfactory both when operated in vacuum cryostat and bare in liquid argon within the \gerda\ setup.
	The measured resolutions at the $Q$-value for double-$\beta$ decay of \Ge (\qbb~= 2039~keV) are about 2.1~keV full width at half maximum in vacuum cryostat. 
	After 18 months of operation within the ultra-low background environment of the GERmanium Detector Array (\gerda) experiment and an accumulated exposure of \expo, the background index after analysis cuts is measured to be $\bidx \times 10^{-4} \ \textrm{counts}/(\textrm{keV} \cdot \textrm{kg} \cdot \textrm{yr})$ around \qbb.
	This work confirms the feasibility of IC detectors for the next-generation experiment \legend.

	\keywords{Double Beta Decay detection \and HPGe well point contact detectors \and Pulse Shape Analysis}
\end{abstract}

\section{Introduction}
As of now the hypothetical neutrinoless double beta (\bb) decay eluded detection in any of the candidate isotopes (e.g.~\Ge~\cite{bib:PRL2020,bib:MjdPRC}, \Te~\cite{bib:CuorePRL}, \Xe~\cite{bib:KamlandPRL,bib:ExoPRL}).
A discovery would have far reaching consequences in particle physics, revealing for the first time lepton number
violation and hence bringing an additional input to the unresolved matter-antimatter asymmetry creation scenarios~\cite{bib:Smirnov2006}.
In absence of a signal at the level of $10^{26}$ yr half-life, next generation experiments are under preparation with significant increase of the isotope mass and further suppression of the background.
Latest $^{76}$Ge-based experiments, GERmanium Detector Array (\gerda) and \mjd, benefited from the high background rejection capability of point contact Ge detectors.
They also feature an excellent energy resolution at the \Ge double-$\beta$ decay $Q$-value (2039~keV), referred to as \qbb\ hereafter, as compared to other detector types.
Among the many challenges inherent to these experiments, one is to scale up the isotope mass by one order of magnitude while lowering the background, which requires a change of germanium detector design.

Previous studies have demonstrated the excellent performance of inverted coaxial (IC) point contact Ge detectors~\cite{bib:Cooper2011,bib:ICPC2018}. 
Their new design allows to increase the detector mass of Broad Energy Ge (BEGe)~\cite{bib:BEGe2015} and P-type, point contact (PPC)~\cite{bib:PPC} detectors from 0.7-1.0~kg up to 2.0-3.0~kg. 
As a result, the amount of surrounding radioactive components like cables and mechanical supports are lowered per detector mass as well as the surface-to-volume ratio. 
The latter lowers the rate of surface background events per mass unit.

The background rejection is based on the analysis of the time profile of the induced current signal, called pulse
shape discrimination (PSD). 
\Ge double-$\beta$ decays in solid germanium feature a single-site energy deposition within $\sim 1$~mm$^3$ (single-site event, SSE). 
Background events from Compton scattered $\gamma$ rays or surface events result in a different pulse shape~\cite{bib:GERDA2013}, as will be discussed below.

\gerda\ operated the germanium detectors in a 64~m$^3$ liquid argon (LAr) cryostat~\cite{bib:cryostat}. 
In May 2018, five enriched IC \Ge detectors, produced in collaboration with Mirion Technologies~\cite{bib:mirion}, have been added to the setup.
The total mass of enriched \Ge detectors has increased by 9.6~kg to 44.6~kg.
Overall they contribute with \expo\ exposure, i.e. 8\% of the total exposure of 103.7~kg$\cdot$yr.
This has been the first time IC detectors have been operated in LAr and important experience has been gained.
PSD together with the veto based on argon scintillation light resulted in the lowest background around \qbb, thus allowing \gerda\ to remain background-free over its entire exposure~\cite{bib:Nature2017,bib:PRL2020}.

The IC detector design has been selected by the \legend\ collaboration, which aims to deploy 200~kg of \Ge detectors in the modified \gerda\ setup by 2021 and ultimately 1~ton in a subsequent stage in order to overpass the $10^{28}$ yr half-life sensitivity~\cite{bib:Abgrall2017}.

In the following, the characterization of the five detectors in vendor vacuum cryostats, performed in March 2018, is described together with a summary of the performance.
Final results within the \gerda\ LAr cryostat after 18 months of operation are shown in the last section.

\begin{figure*}
	\centering
	\includegraphics[width=0.367\linewidth]{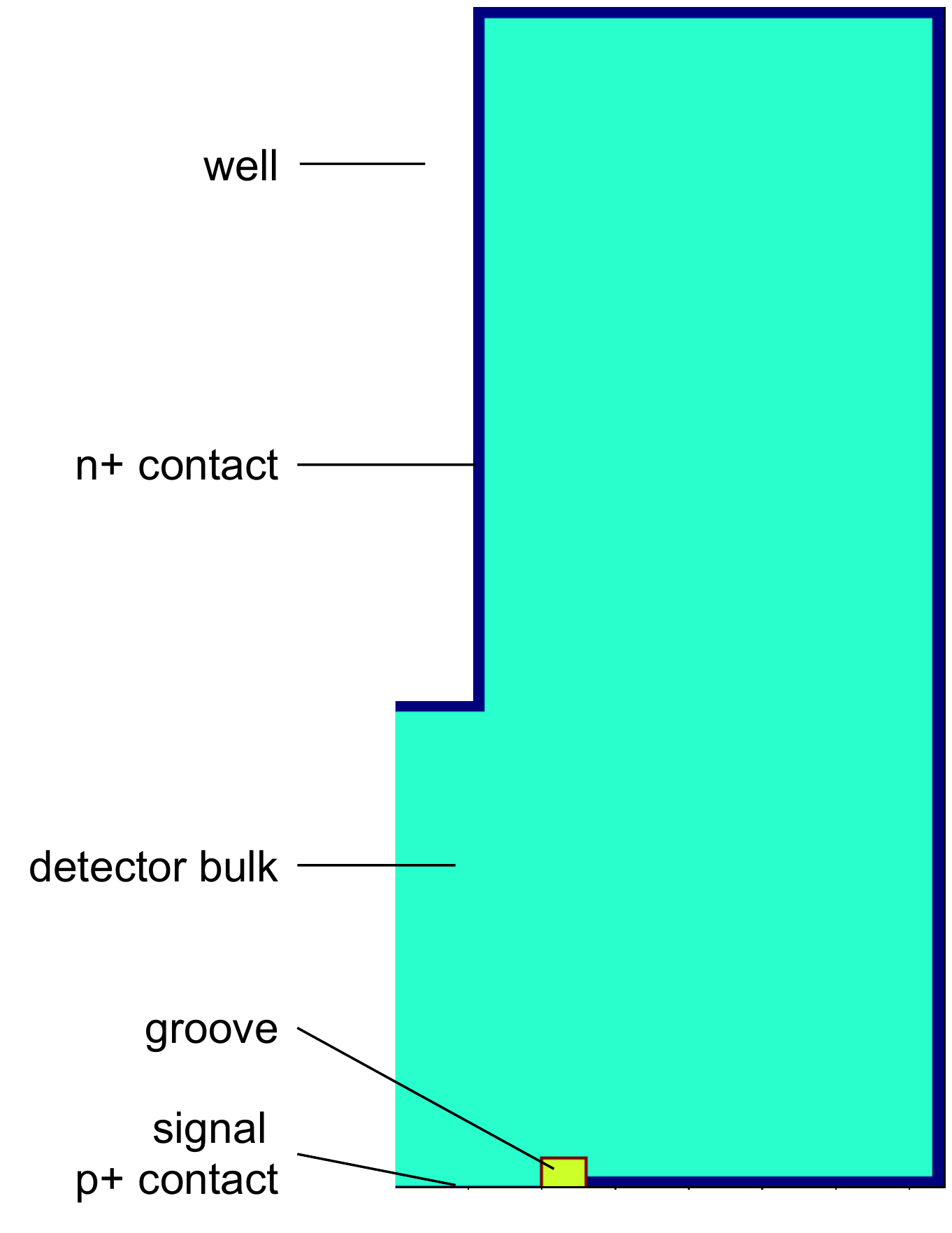} \hfill
	\includegraphics[width=0.274\linewidth]{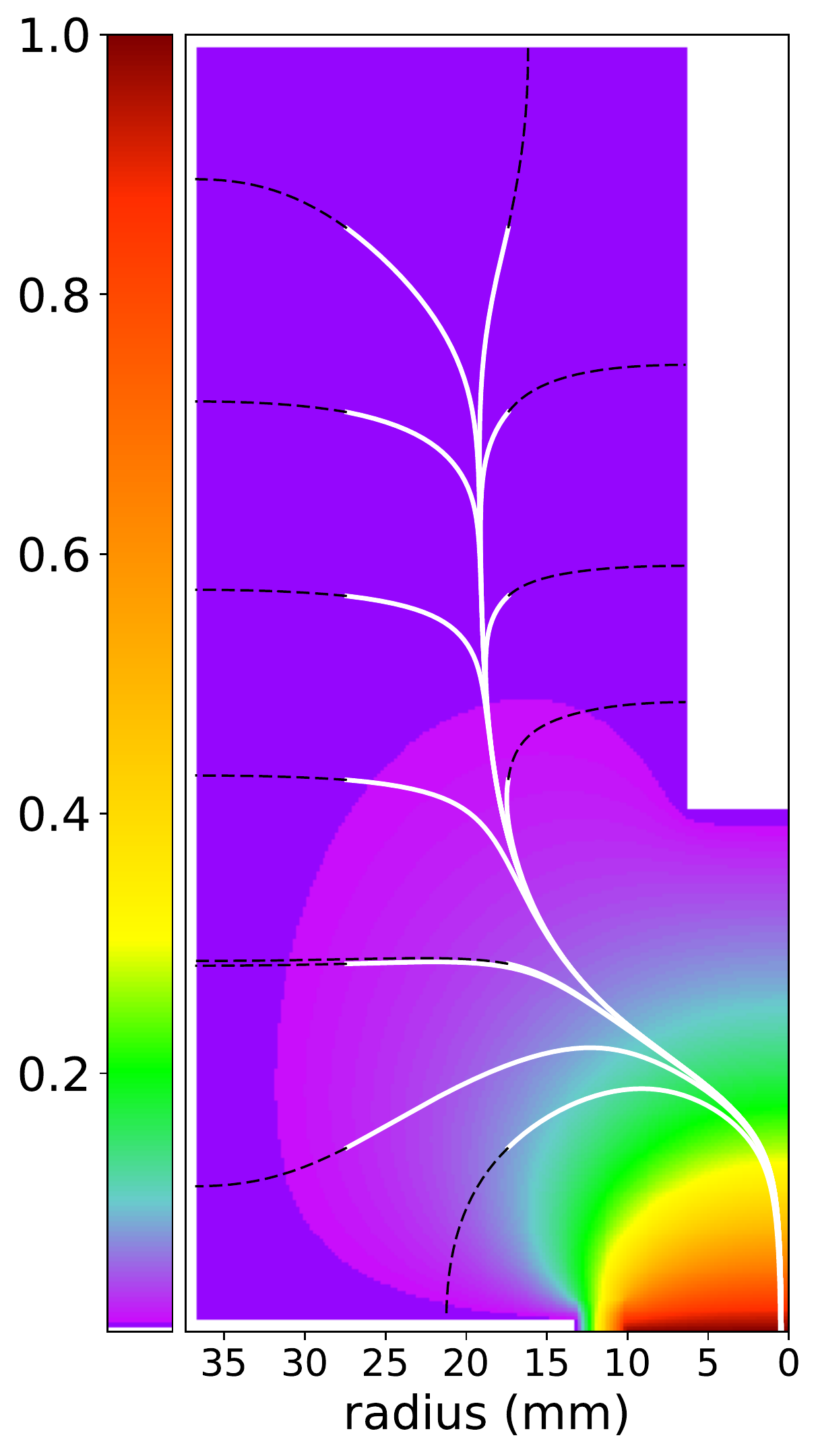}
	\includegraphics[width=0.303\linewidth]{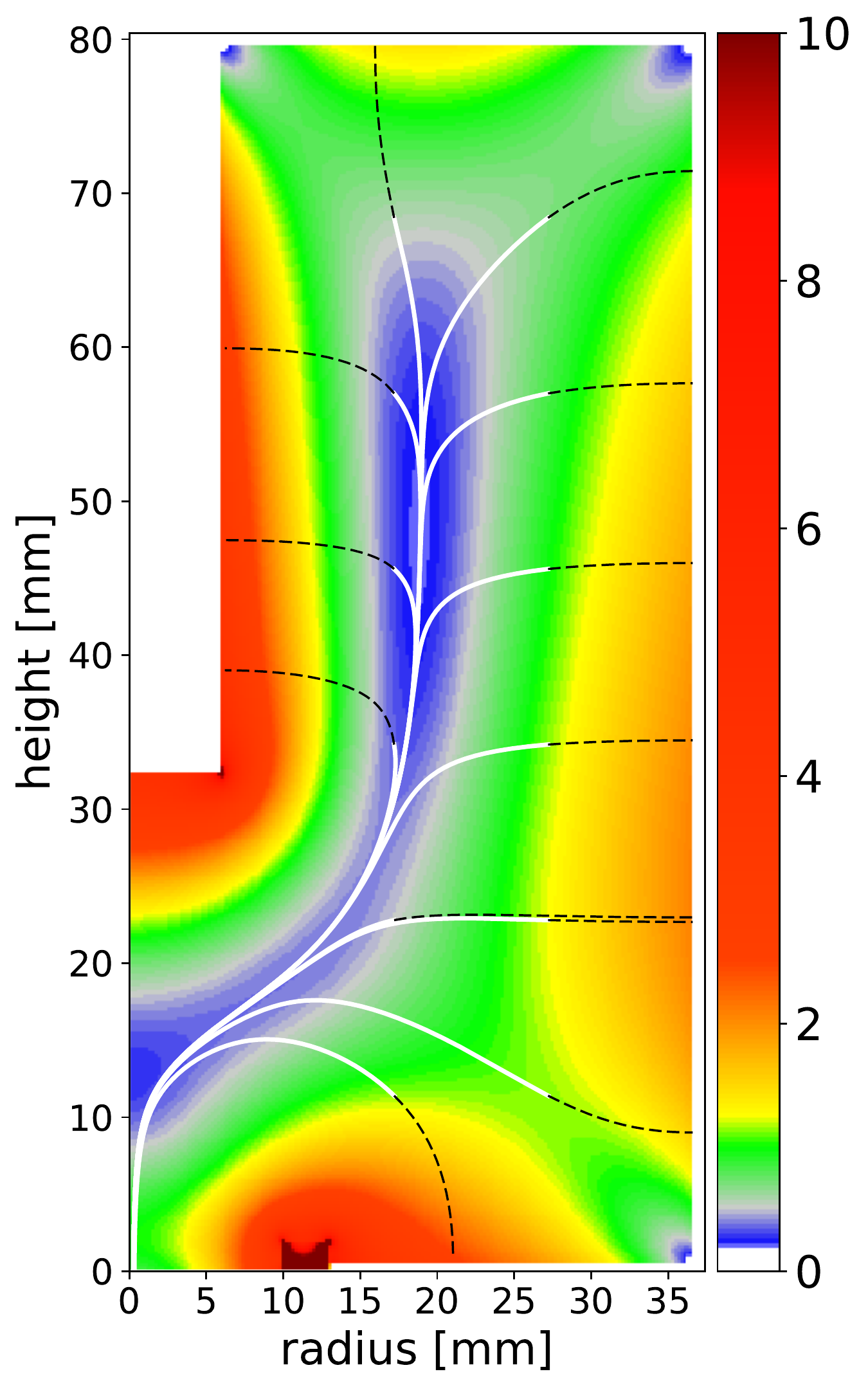}
	\caption{Left: Main IC detector features.
		Middle: ADL calculation of the weighting potential. 
		Right: Electric field strength in kV/cm.
		The minimum required electric field is 200~V/cm (dark blue).
		The black dashed lines show electron drift paths ending at the n+ contact while white solid lines are the hole drift paths reaching the signal contact.}
	\label{fig:IC_design}
\end{figure*}

\begin{figure*}
	\centering
	\includegraphics[scale=0.5]{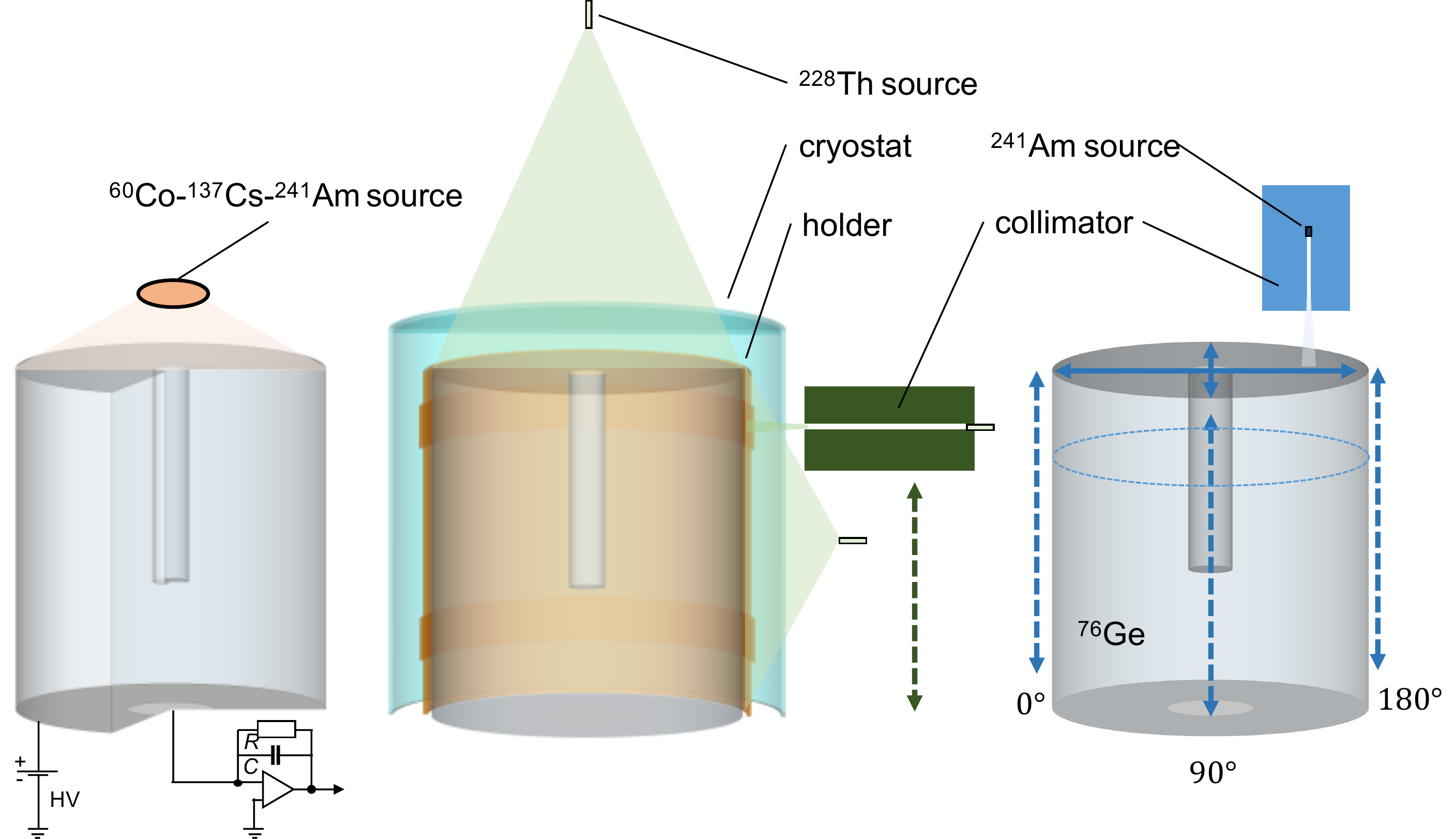}
	\caption{Configurations used for detector characterization.
		Left: Setup for depletion voltage estimation with a mixed 1.5 kBq source of \Co-\Cs-$^{241}$Am; bias and readout circuits are indicated.
		Middle: Setup for PSD studies with a flood top and side 13 kBq \Th source and with a collimated 250 kBq \Th source for lateral scans.
		The vacuum cryostat (cyan) and detector holder (orange), both made of aluminum, are added here for illustration.
		Right: Setup for scans with the collimated 4.3 MBq \Am source: lateral at 3 azimuthal angles (dashed lines), 2 orthogonal directions on top (solid lines) and a circular one (dotted lines).}
	\label{fig:setup}
\end{figure*}

\section{Inverted Coaxial design and detector production}
\label{sec:Production}

In 2017, 20~kg of germanium, enriched at 87.7(5)\% in $^{76}$Ge, have been purchased in form of GeO$_2$ powder.
The reduction and purification process to convert the GeO$_2$ to 50 $\mathrm{\Omega}\cdot$cm electronics grade \Ge material took place at the PPM Pure Metals company in Goslar, Germany. 
About 95\% of the material had the required quality and 1.6\% was lost during etching and cutting of the bars. 
The rest of the material was too little for another purification process (zone-refinement) and has been stored for future processing.

The \Ge material bars were shipped to Mirion Technologies -- Oak Ridge, US in order to further proceed with crystal pulling.
Three large crystals were grown and cut into 5 slices with a total mass of 9.826~kg.
A loss of 1.5~kg from etching, cutting and grinding has been reported at this stage.
Another 4~kg of \Ge material with unsatisfactory purity or remaining of cutting and grinding processes will enter a future production cycle after chemical purification.
The final detector fabrication stage took place in February 2018 for about two weeks at Mirion Technologies -- Olen, Belgium.
A total of 9.618~kg were converted into 5 working IC detectors.
Overall, the estimated production mass yield was 51\%, slightly below the achieved 53.3\% during the previous BEGe detectors production~\cite{bib:BEGe2015}.

During production, care was taken to minimize the detector exposure to cosmic rays.
It meant storing detectors in underground places near manufacturer sites and transporting them across ocean in a shielded container~\cite{bib:BEGe2015}.
A total of 36 days-equivalent time above ground over one year has been estimated.

\begin{figure}
	\centering
	\includegraphics[width=\linewidth]{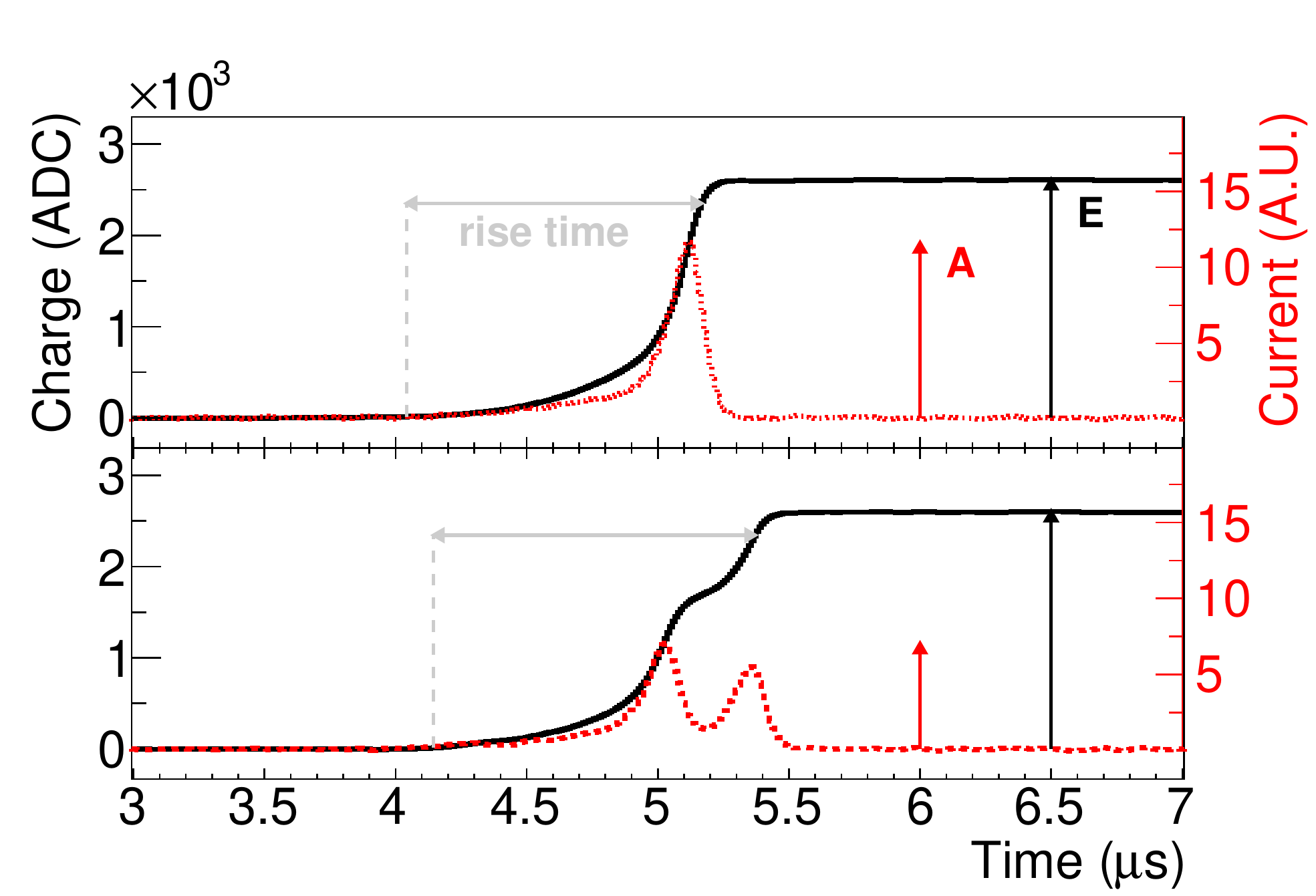}
	\caption{Waveform examples of SSEs (top) and MSEs (bottom) of detector 50A after 
	applying a moving window average.
	The amplitude of the maximum current $A$ and of energy $E$ are explicitly shown.
	The boundaries of the rise time are estimated at 0.5\% and 90\% of the maximum charge amplitude.}
	\label{fig:Pulses}
\end{figure}

In Fig. \ref{fig:IC_design}(left), the geometry of an IC detector is sketched. 
The detector is cylindrically symmetric.
The important novelty of the IC design is the well opposite to the p+ contact. 
The electric field (see Fig.~\ref{fig:IC_design}(right)) created by the bulk net impurities and the externally applied high voltage pushes holes towards the p+ contact and electrons towards the n+ contact. 
The well is also important to deplete the detector at the usual operational voltage of about 4000~V.
Detectors without the well (e.g. BEGes) are limited to less than 1~kg of mass while IC detectors might have a mass above 3~kg.
Typical charge collection times of such detectors lie within 1 $\mu$s to 1.5 $\mu$s (see Fig. \ref{fig:Pulses}) which is 2 to 3 times larger than that of BEGes.

Given an impurity concentration profile of the bare Ge crystal from the manufacturer after the growth, the geometrical dimensions were optimized via the electrostatic simulation package ADL~\cite{bib:Bruyneel2016} by \gerda\ in order to ensure a minimal electric field of 200~V/cm and a maximum depletion voltage of 4000~V.
Based on this optimization, the detectors were finally machined by the manufacturer.

The current signal $I(t)$ induced by the drifting charges $q$ as a function of time $t$ can be calculated with the weighting potential $\Phi(\vec{r})$ according to the Shockley-Ramo theorem~\cite{bib:He2001}.
\begin{equation}
	I(t) = q \cdot \nabla \Phi(\vec{r}(t)) \cdot \vec{v}(t)
\end{equation}
with $\vec{v}$ being the drift velocity of the charges. 
Fig.~\ref{fig:IC_design}(middle) shows the weighting potential. 
$I(t)$ is maximal in the vicinity of the p+ contact. 
Since all holes follow the same path at the end of their drift, close to the p+ contact (funnel effect \cite{bib:BEGe2009}), the maximum $A$ of the induced current is independent of the starting point. 
As a consequence $A/E$, with $E$ being the deposited energy, is constant for all SSEs, while it will be reduced for multiple well-separated depositions (multi-site events, MSEs - see Fig. \ref{fig:Pulses}) and reduced or enhanced for surface events. 
Latter events, originating in \gerda\ from $^{42}$K $\beta$ decays on the n+ contact 
and $\alpha$ particles on the p+ contact \cite{bib:Bkg2020}, are not studied in this paper due to the inadequate setup.
The PSD for such point contact detectors is therefore primarily based on the ratio $A/E$~\cite{bib:GERDA2013,bib:MJDpsd}.

In order to understand the response of the new large detectors and evaluate the effect of this well on the detector performance, an extensive characterization with radioactive sources was needed.

\section{Detector characterization in vacuum cryostat}
\label{sec:Characterization}

The characterization took place in the HADES (High Activity Disposal Experimental Site) underground laboratory in Mol, Belgium~\cite{bib:HADES}, benefiting from a 500 m water equivalent overburden with negligible hadronic background and low \Ge activation rate from atmospheric muons.
A description of the HEROICA platform that was used can be found in~\cite{bib:HEROICA}. 
Similar measurements were performed and the same analysis protocols were applied as in~\cite{bib:BEGe2019}.
A summary of the general detector information is given in Table \ref{tab:Det_infos}.

\begin{table}[h!]
	\centering
	\caption{Diameter ($D$), height ($H$), well depth ($W$) and mass ($M$) of the five inverted coaxial detectors.
		Their well diameter is 10.5~mm.
		The dimensions were provided by the manufacturer while masses were measured by \gerda.}
	\begin{tabular}{c|ccccc} 
		Det. ID & 48A & 48B & 50A & 50B & 74A  \\ \hline
		$D \pm 0.2$ [mm] & 74.6 & 72.6 & 74.2 & 72.6 & 76.6  \\ 
		$H \pm 0.3$ [mm] & 80.4 & 80.5 & 80.4 & 85.4 & 82.3  \\
		$W \pm 0.5$ [mm] & 47.4 & 56.0 & 40.0 & 53.9 & 52.4  \\
		$M \pm 0.5$ [g] & 1918.9 & 1815.8 & 1881.1 & 1928.7 & 2072.9
	\end{tabular}
	\label{tab:Det_infos}
\end{table}

\subsection{Experimental setup}
\label{sec:Setup}

Each detector was mounted in a vacuum cryostat and cooled down to 95~K by a cold finger immersed in liquid nitrogen.
In total, three test benches were used, two static tables with lead and copper shields and one unshielded scanning table (see Fig. \ref{fig:setup}), with the following goals.
\begin{itemize}
	\item \textbf{Determine the nominal bias voltage}. The analysis of the 1333~keV $\gamma$ line properties of \Co\ was performed for various applied bias voltages. 
	\item \textbf{Probe the geometrical detector response}. The homogeneity of the detectors' surface response was finely scanned by embedding a highly collimated \Am source on a 3D movable arm~\cite{bib:HEROICA}.
	The source was typically moved by steps of 1~mm every 5~min.
	The 3 cm thick collimator with an aperture of 1~mm diameter produces a $\sim 2\ \textrm{mm}^2$ spot for which 95\% of the 60~keV $\gamma$ rays are absorbed in \Ge within 3~mm.
	\item \textbf{Estimate the best achievable energy resolution}. Construct the resolution curve from all available $\gamma$ lines full width at half maximum (FWHM) as measured in optimal electronic noise data taking condition.
	\item \textbf{Evaluate the pulse shape discrimination performance}. The \Th energy spectrum and pulse shape parameters were recorded for typically a few hours using a 12~kBq non-collimated source to probe the overall detector bulk response. 
	Additionally, due to the large size of the IC detectors, these parameters have been coarsely scanned at four different heights with a 250~kBq \Th source mounted behind a 20~cm long collimator of 10~cm diameter with an aperture of 2.5~mm diameter.
	This special measurement was performed at the Technische Universit\"at M\"unchen underground laboratory.
\end{itemize}

A \gerda-like data acquisition system~\cite{bib:upgrade} was used for these measurements.
The signal was first amplified before being digitized at 100~MHz by a VME-Struck FlashADC (FADC).
For each triggered event, a 10~$\mu$s long trace and a 25~MHz down-sampled trace of 160~$\mu$s length were stored to disk for offline processing.
The former trace was used to study the shape of the current pulse, i.e. $A$ and signal rise time, while the latter was used for the energy reconstruction. 
The standard \gerda\ analysis software, \textsc{Gelatio}~\cite{bib:GELATIO} has been used for this study.
For the energy reconstruction, a trapezoidal digital filter has been selected with the following parameters: 4~$\mu$s rise and fall time and 2~$\mu$s flat top.
Due to occasional FADC failures, part of the data have been recorded by a Multi-Channel Analyser (MCA).
This acquisition mode provided an energy spectrum measured at 10$~\mu$s shaping time without pulse shape information.

\subsection{Nominal bias voltage}
\label{subsec:DV}

\begin{figure}
	\centering
	\includegraphics[width=\linewidth]{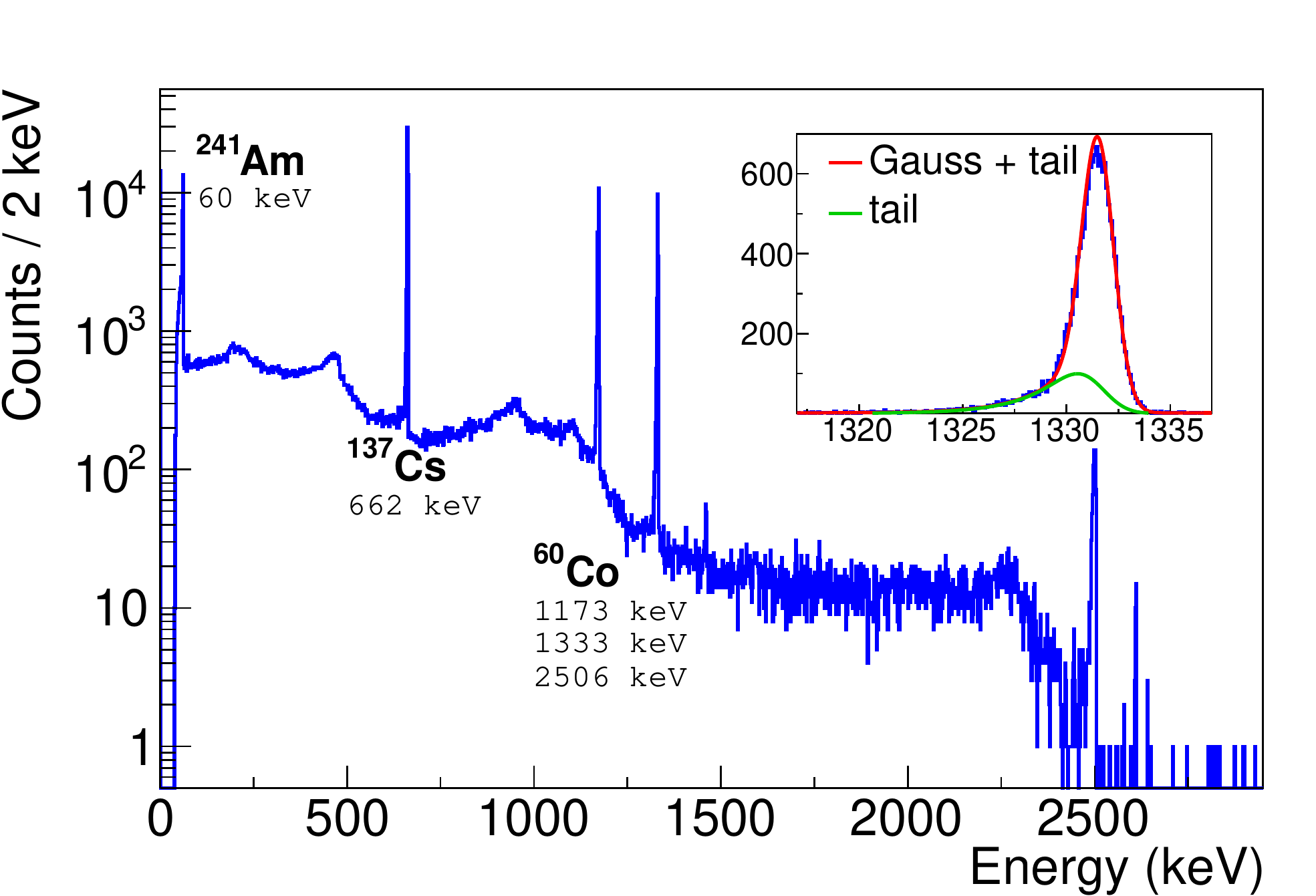}
	\caption{Spectrum taken with detector 50A and the mixed source of \Co, \Cs, \Am\ for the determination of the nominal bias voltage.
		The inset shows the fit to the \Co\ 1333~keV $\gamma$ line.
		The MCA module was used for this measurement using a Gaussian energy filter, thus explaining the significant tail of the $\gamma$ line from ballistic deficit.
	}
	\label{fig:HV_scan_spectrum}
\end{figure}	

\begin{figure}
	\centering
	\includegraphics[width=\linewidth]{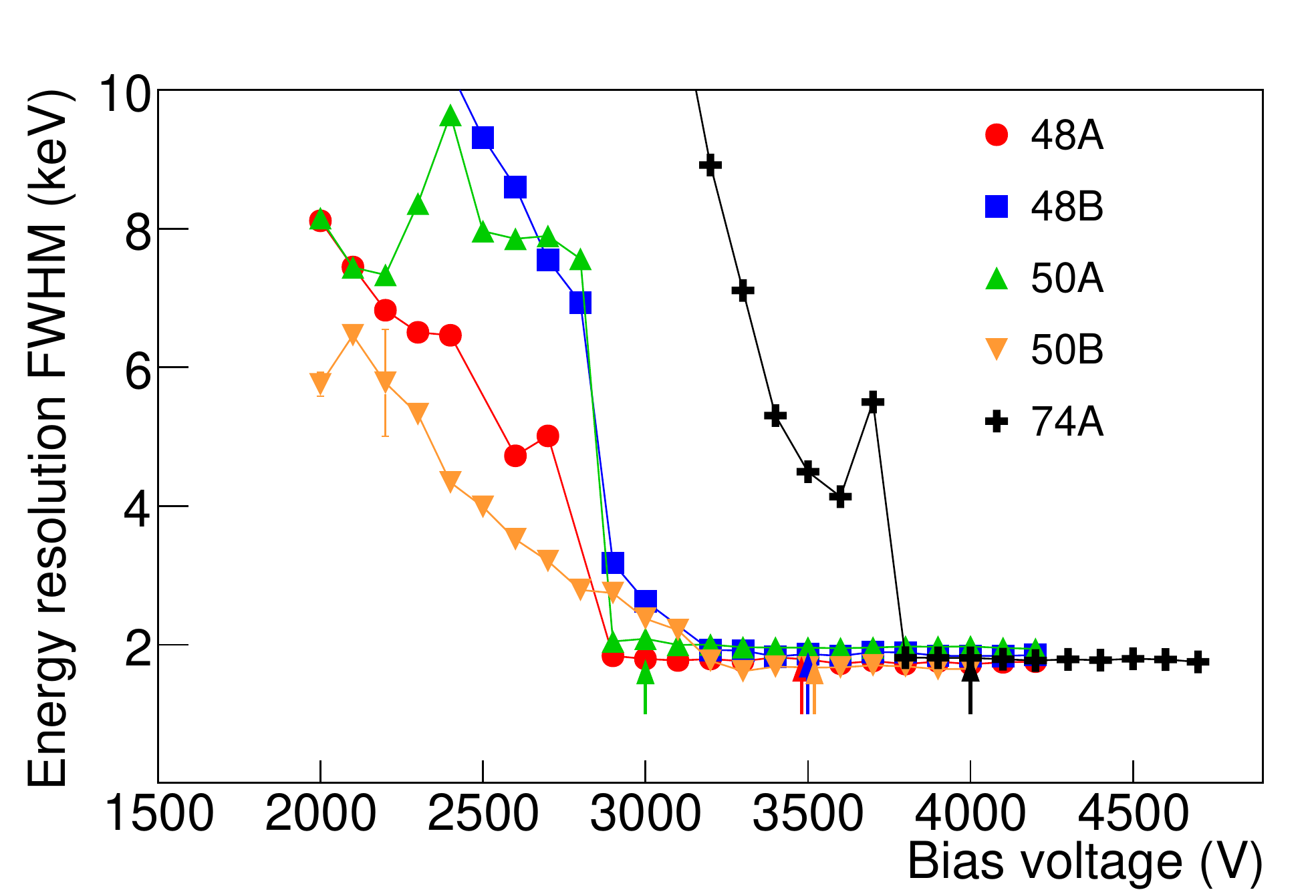}
	\caption{Energy resolution FWHM of the \Co\ 1333~keV line as a function of the applied bias voltage.
	The arrows show depletion voltages and resolutions reported by the manufacturer.
	The statistical uncertainties are less than the widths of the markers.
	}
	\label{fig:HV_scan}
\end{figure}

The energy resolution of the detectors has been studied under various applied bias voltages in order to retrieve the nominal bias voltage.
It is defined here as the voltage where the energy resolution reaches its minimal and stable value.
The bias voltage was varied from 2000~V up to 4700~V in steps of 100~V and data taken with a mixed 1.5~kBq source of \Co-\Cs-\Am (see Fig. \ref{fig:HV_scan_spectrum}).
The results of the five detectors are illustrated for the 1333~keV \Co\ line in Fig. \ref{fig:HV_scan} and the retained nominal bias voltage for each detector is listed in Table \ref{tab:Dep_results}.
The peak position and peak integral variations were also studied. For each detector, the deduced 
nominal bias voltage agreed within 100~V with the estimate from the energy resolution 
measurements.
All nominal bias voltages were found to be lower than 4000~V, fulfilling the required specification.

Theoretical field calculations were performed with the ADL software~\cite{bib:Bruyneel2016} for gradually increasing bias voltages. 
Full depletion was assumed when a minimal electric field of 10 V/cm was reached everywhere in the detector bulk. 
The deduced depletion voltages agreed with the data at the level of 100 V, except for detector 50A (see below).

The disagreement can mainly be attributed to the uncertainty on the net impurity concentration profile (amplitude and gradient) provided by the manufacturer.
As a general rule, the operational voltage in the \gerda\ experiment was set to 400-600~V above the quoted nominal bias voltage (see Table \ref{tab:Dep_results}).
Detector 50A exhibits a peculiar behaviour: different from the other detectors, its energy resolution remains rather constant between 2000~V and 2800~V, and the ADL simulation predicts a nominal bias voltage that is lower by 760~V than measured. 
The ADL simulation does explain these observations showing a pinch-off 
(see discussion in \cite{bib:BEGe2019}) at bias voltages within 2100~V to 2700~V, 
i.e. a persisting very low electric field region close to the p+ contact. 
Hence a much higher bias voltage is needed to establish sufficient field strength at the p+ contact.

As to the recommended operational voltage, we verified that the norm of the electric field 
fulfilled the minimal electric field specification of 200~V/cm (see Sect. \ref{sec:Production}).

Detector 48A could not be operated beyond the nominal bias voltage in \gerda\ because the noise from the strongly increased leakage current led to unacceptable data quality.

For characterization measurements in vacuum cryostats, described in the following, 
the voltages recommended by the manufacturer, which are typically 500~V to 1000~V higher than 
the depletion voltages, were applied (see Table \ref{tab:Dep_results}).

\begin{table}[h!]
	\centering
	\caption{Nominal bias voltages from scans compared to simulations and manufacturer 
	         depletion voltages. 
		Also shown are the operational voltages of the measurements at HADES and \gerda.}
	\begin{tabular}{r|ccccc} 
		\multicolumn{1}{c|}{Det. ID} & 48A   & 48B   & 50A   & 50B   & 74A  \\ \hline
		Nominal volt. [V]  	& 2900 & 3200 & 2900 & 3200 & 3800  \\ 
		ADL simu. [V]  		 & 3010 & 3335 & 2140 & 3320 & 3620  \\
		Manufacturer [V]   & 3500 & 3500 & 3000 & 3500 & 4000 \\ \hline
		Set in HADES [V]   & 4000 & 4000 & 4000 & 4000 & 4500 \\
		Set in \gerda\ [V]	 & 3300 & 3200 & 3700 & 3800 & 4400 \\
	\end{tabular}

	\label{tab:Dep_results}
\end{table}

\subsection{Energy resolution}
\label{subsec:Eres}

\begin{figure}
	\centering
	\includegraphics[width=\linewidth]{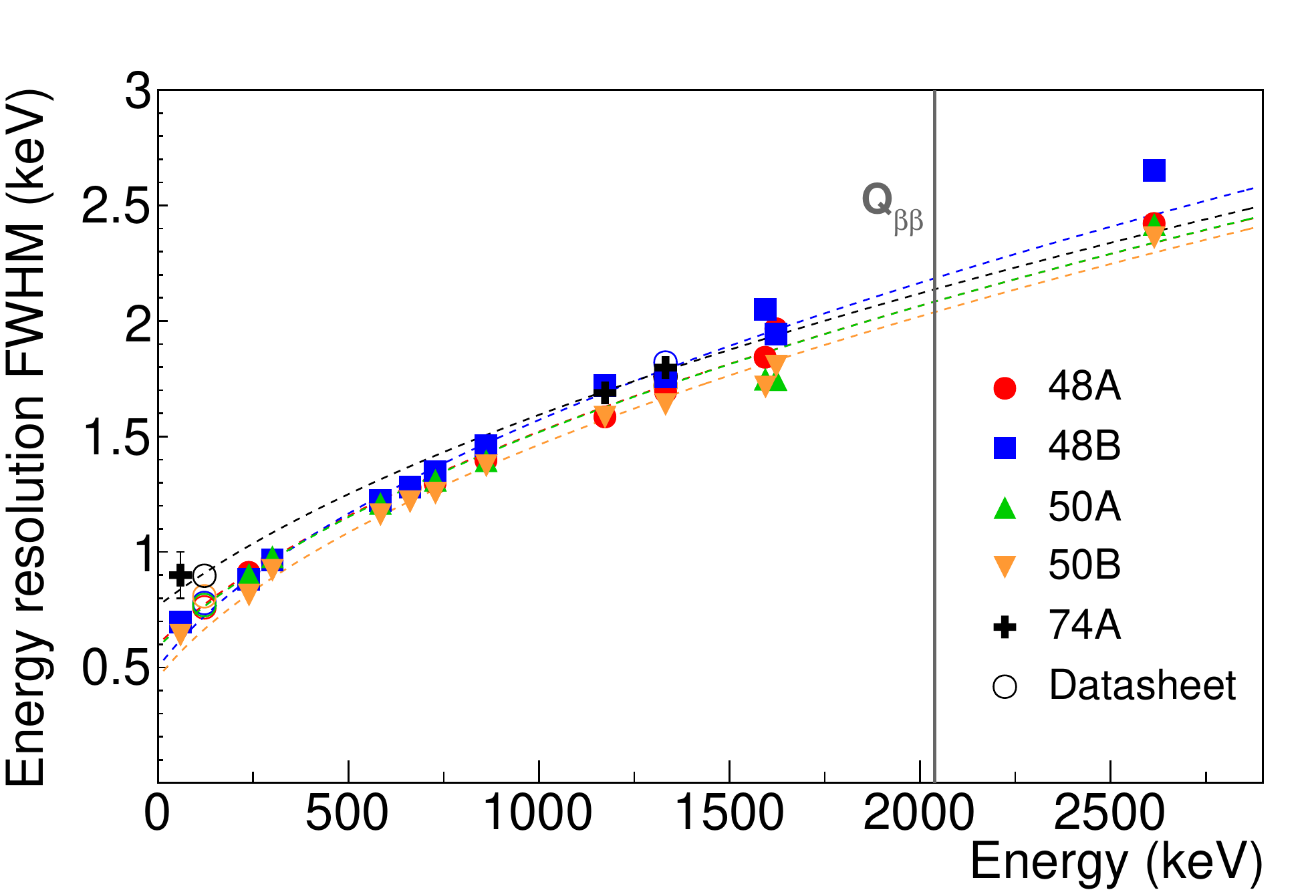}
	\caption{Energy resolution FWHM as a function of $\gamma$ ray energy; data are taken at the bias voltage recommended by the manufacturer. 
		The dashed lines show fits to the data, performed for each detector separately.
		The statistical uncertainties are less than the widths of the markers.}
	\label{fig:Eres}
\end{figure}

The energy resolution of all available prominent $\gamma$ lines from the full data collection (cf. Fig. \ref{fig:setup}) is shown for each detector in Fig. \ref{fig:Eres}.
The FWHM resolution is estimated from a Gaussian fit to the data after subtraction of the background observed in the side bands.
Results compatible with the specification from the manufacturer were found.
A global fit ($\sqrt{a + b \times E}$) to the data is used to interpolate the energy resolution at $Q_{\beta\beta}$ which ranges from 2.04~keV to 2.19~keV FWHM in vacuum cryostat.

\subsection{Surface response homogeneity}
\label{subsec:SurfaceScan}

\begin{figure}
	\includegraphics[width=\linewidth]{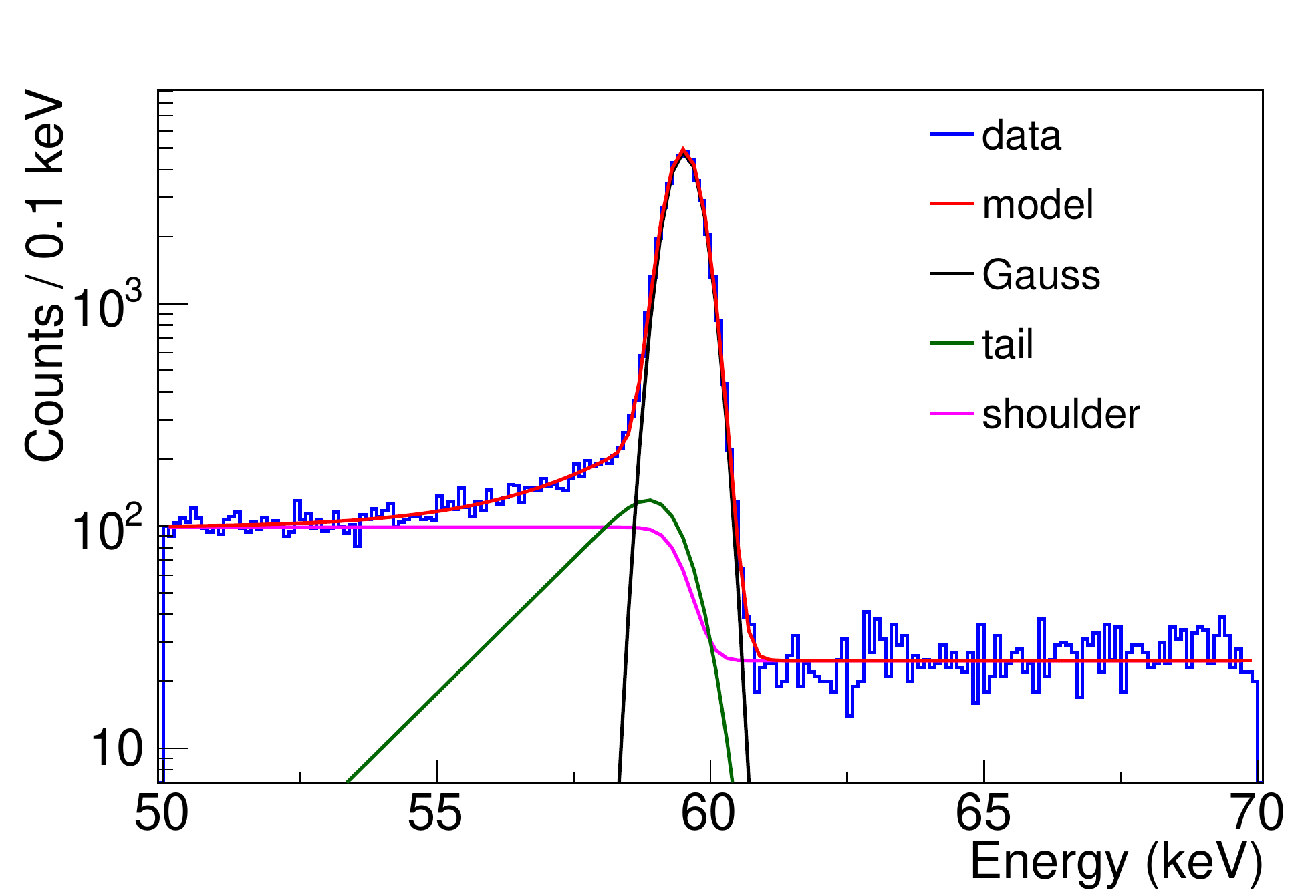}
	\caption{Example of \Am FEP data recorded with the highly collimated 4.3 MBq source, positioned above the upper surface of detector 50A.
		The model (red line) is shown together with its decomposition into a Gaussian, a tail and a shoulder functions.}
	\label{fig:AmFit}
\end{figure}

\begin{figure*}
	\includegraphics[width=\linewidth]{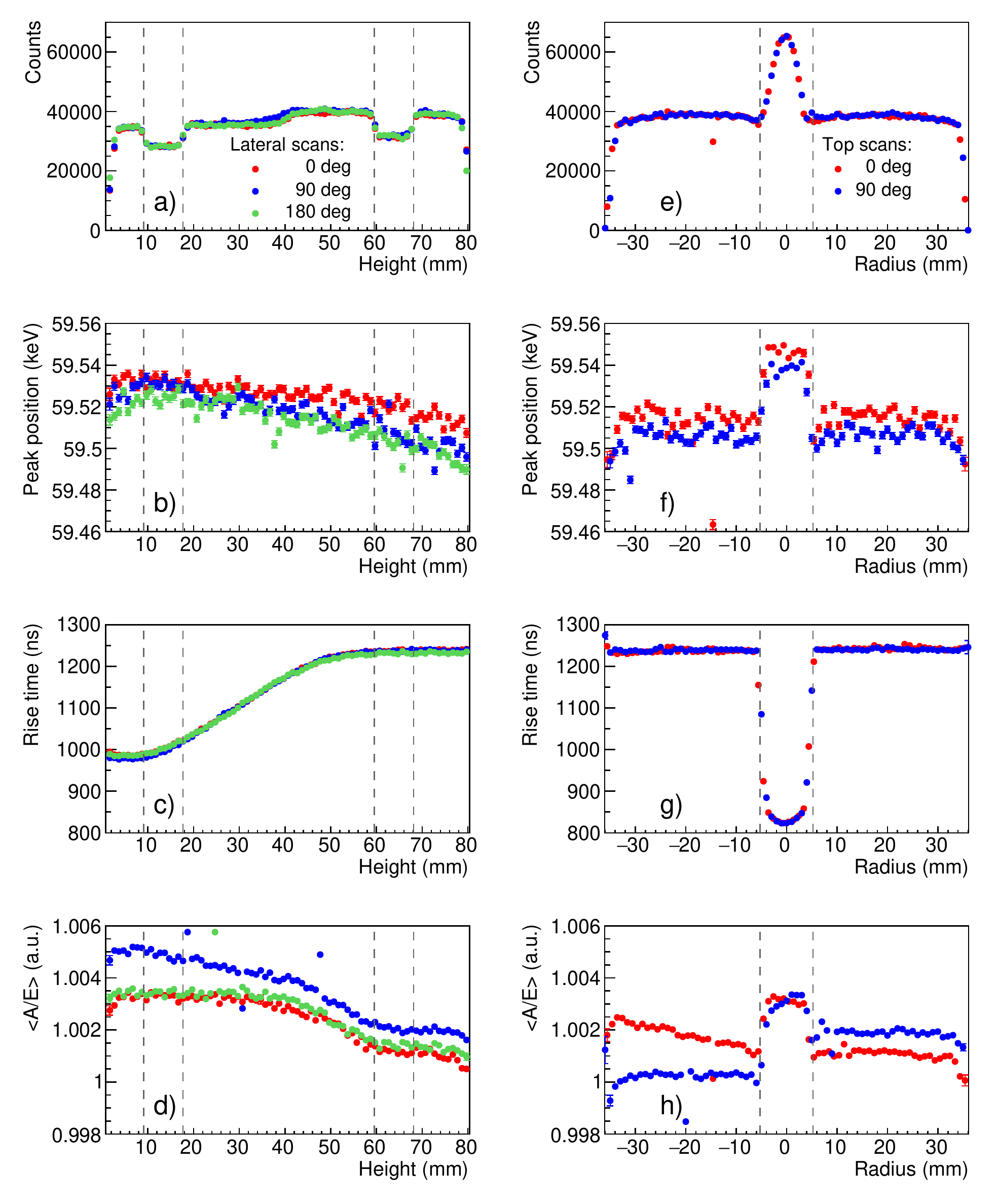}
	\caption{Results of the lateral (left) and top radial (right) scans for the 60~keV \Am $\gamma$ line obtained with detector 50A. 
		Statistical uncertainties are less than the widths of the markers.
		The dashed gray lines on the left(right) show the expected holder ditch (well) position (see Fig. \ref{fig:setup}).}
	\label{fig:AmScan}
\end{figure*}

In this section, detector response parameters of interest for the data analysis in \gerda\ are detailed as a function of the interaction position near the detector surface.
We investigated the charge collection efficiency, the [0.5\%--90\%] signal rise time and the pulse shape discrimination parameter by means of scans with a highly collimated low energy $\gamma$ ray source (60~keV $\gamma$ line from a 4.3 MBq \Am source).
In total, about 1500 measurements were performed with an exposure of 5~minutes each.
For the sake of conciseness, we show in Fig. \ref{fig:AmScan} and comment below a selection of relevant observables for detector 50A and events originating from the prominent 60~keV line.
The number of counts and the peak position are obtained after Compton background and tail subtraction (see Fig. \ref{fig:AmFit}).
Events found in the tail are typically from $\gamma$ rays with partial charge collection due to energy loss at the detector surface~\cite{bib:BEGe2019,bib:MA2017}.

Firstly, the lateral scans (Fig. \ref{fig:AmScan}a) feature a specific count rate profile mainly affected by the non homogeneous width of the detector holder (cf. two 9~mm ditches in Fig \ref{fig:setup}, middle).
The profile is well reproduced at different angles.
This specific detector has a slight taper on the upper part at heights above 40~mm, leading to a thinner full charge collection depth (FCCD) in this region hence explaining the visible higher count rate starting at $H=40$~mm.
The charge collection efficiency of the 60~keV $\gamma$ line, quantified via the peak position (Fig. \ref{fig:AmScan}b), decreases as the energy deposition occurs further away from the p+ contact.
This behavior, observed at all angles and for all detectors, is understood as charge trapping along the drift path.
The effect is lower ($\sim0.05$ \%) than the energy resolution at \qbb\ (0.13\%) but sizable.
Thanks to its correlation with the charge collection time, it can effectively be corrected for.
The reported rise time [0.5\%--90\%] is the average value of all the events in the peak.
Its mean distribution (Fig. \ref{fig:AmScan}c) increases linearly in the lower part of the detector before reaching a plateau.
This plateau is best explained by the charge drift paths (see Fig. \ref{fig:IC_design} right), all converging at a height of 5~cm toward the same low electric field region.
Finally, the normalized pulse shape parameter $\left<A/E\right>$ is also scrutinized.
$\left<A/E\right>$ is the position of the Gaussian peak of the $A/E$ distribution of 60~keV events.
As seen in Fig. \ref{fig:AmScan}d) it features two plateaus that are understood via the simulation of charge cloud dynamics.
This dynamics comprises diffusion, acceleration and repulsion of the charges that, when summed, increase the spatial distribution of the charge clouds as they start drifting towards the p+ contact.
As a consequence of a larger spatial distribution, a lower current amplitude $A$ for the upper scan positions is observed.
Electrostatic simulations, performed with the Siggen software~\cite{bib:Siggen}, reproduce this feature both qualitatively and quantitatively, as detailed in~\cite{bib:Comellato2020}.

Secondly, the top radial scans shown in Fig. \ref{fig:AmScan}, right, also have an interesting pattern due to the presence of the well.
While the average count rate is in agreement with the upper lateral measurement, an increase at the well location is explained by both an increased acceptance and a thinner FCCD originating from the fabrication process.
Same observations apply to the charge collection.
The mean rise time profile is stable at 1240~ns but peaks off to 820~ns in the well, close to the p+ contact.
The $\left<A/E\right>$ shift observed at the well level confirms the interpretation given for the lateral scan since no difference in $\left<A/E\right>$ between well and top events is expected from pure electrostatic simulation, without charge cloud dynamics\footnote{Energy depositions in the vicinity of the p+ contact do have a significantly higher $\left<A/E\right>$ due to higher weighting field gradient but only within a $\sim$1 cm distance from the contact.}.
The small angular $\left<A/E\right>$ shifts apart from the well can originate from a slight misalignment of the well or the groove ($<0.5$~mm) relative to the central detector axis that would create an asymmetric weighting field.

The outcome of these exhaustive scans is summarized for each detector in Table \ref{tab:AmScan} in form of the observed maximal variation of each parameter between the various source positions.
From this Table, one can appreciate that the shift of the 60~keV $\gamma$ line for the entire data collection lies within the same order of magnitude of the energy resolution at $Q_{\beta\beta}$, 0.1\% and 0.15\% in vacuum cryostat and in LAr (see Sect. \ref{sec:Operation}), respectively.
As a result, despite the large drift time, a sufficiently good homogeneity of the charge collection over the whole detector volume is ensured.
Concerning $\left<A/E\right>$, shifts of its average value vary from 0.2\% up to 1.0\%.
This is of the same order of magnitude as the typical $A/E$ resolution found at higher energies, close to $Q_{\beta\beta}$ in vacuum cryostat (see Fig. \ref{fig:AoE_scan}).
However, the shift is much lower than what was achievable in \gerda\footnote{The $A/E$ resolution is of the order of 1.8\% in the \gerda\ LAr cryostat due to much worse electronics noise.}~\cite{bib:GERDA2013}.
The minimal and maximal rise times are taken from the lateral measurement, i.e. the most relevant values for benchmarking \Th source data (see Sect. \ref{subsec:PSD}).
For detector 50B, only MCA data were taken, hence no pulse shape parameters are available and the peak position is less reliable.

\begin{table}[h!]
	\centering
	\caption{Maximal variation $\Delta$  of peak position (PP) and $\left<A/E\right>$ at the 60~keV $\gamma$ line.
	The minimum and maximum rise time (RT) of the 60~keV $\gamma$ line is reported for comparison with \Th\ data.}
	\begin{tabular}{c|ccccc} 
		Det. ID 			  & 48A  & 48B   & 50A  & 50B   & 74A  \\ \hline
		$\Delta$PP (\%)  & 0.06 & 0.20  & 0.08  & 0.33  & 0.13  \\ 
		$\Delta \left<A/E\right>$ (\%)& 0.21	& 1.05  & 0.43  & -       & 0.75 \\ 	
		min RT (ns)  	   & 970   & 925   & 980   & -       & 920 \\ 
		max RT (ns) 	  & 1190 & 1200  & 1240 & -       & 1150 \\ 	
	\end{tabular}
	\label{tab:AmScan}
\end{table}

\subsection{Pulse shape discrimination performance}
\label{subsec:PSD}

\begin{figure}
	\centering
	\includegraphics[width=\linewidth]{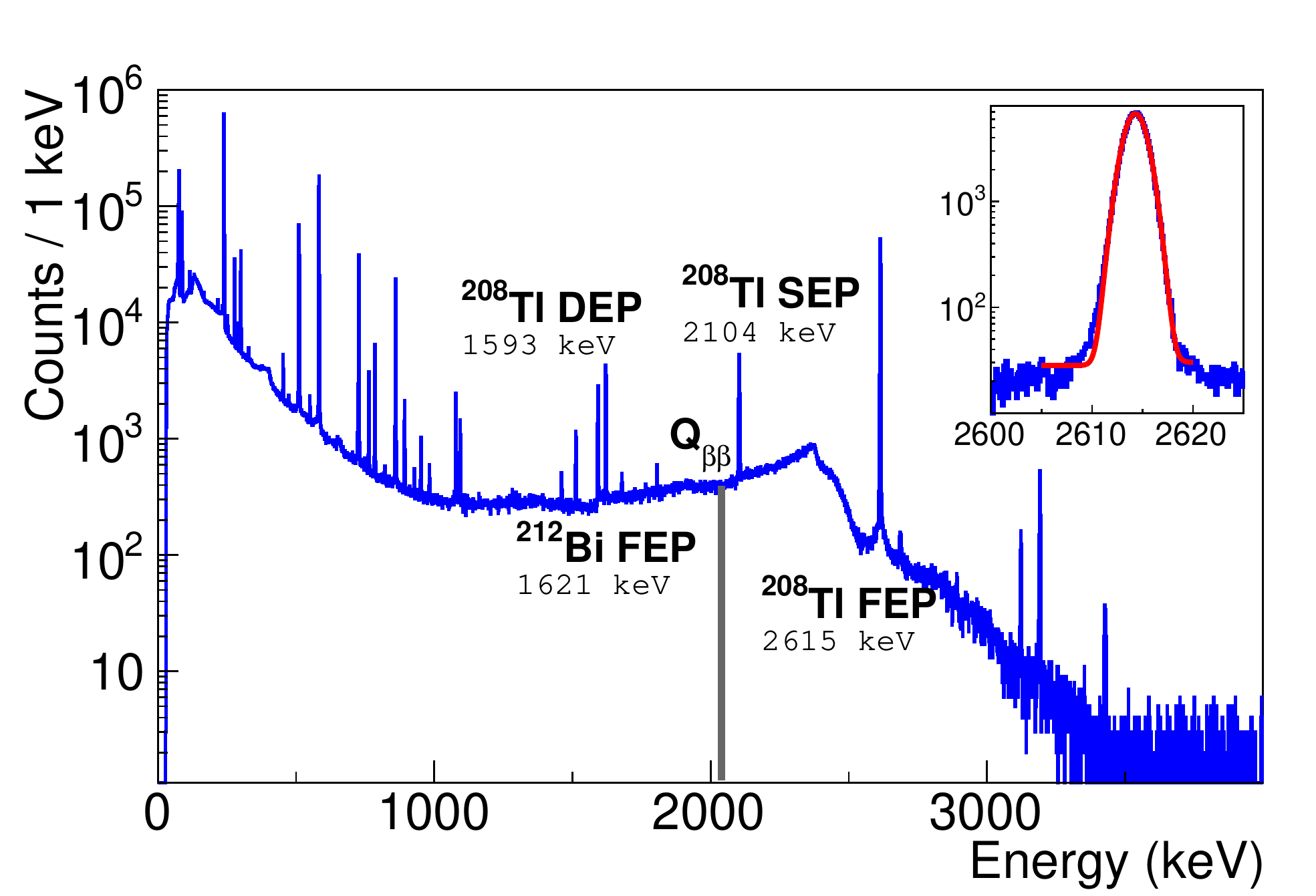}
	\caption{\Th spectrum taken with detector 50A and the source in lateral position.
		The main $\gamma$ lines, \Tl double escape (DEP), \Tl single escape (SEP) and \Tl \& \Bi full energy (FEP) peaks, used in the pulse analysis, are emphasized together with the \qbb$\pm 35$~keV Compton continuum region.
		The inset shows a fit to the 2615~keV $\gamma$ line.}
	\label{fig:Th_spectrum}
\end{figure}

Pulse-shape analysis provides a simple but powerful technique to discriminate signal-like SSEs from background-like MSEs which is crucial for realizing the background-free condition~\cite{bib:Nature2017}.
In the following analysis, the \Tl double escape peak (DEP) has been used as a proxy for SSEs while MSEs have been considered from various Compton scattering regions, including the Compton continuum at \qbb$~\pm~35$~keV, referred to as CC~@~\qbb.
During the characterization campaign, three types of data were recorded with a \Th source: two flood measurements from top and lateral positions and collimated lateral measurements at four different heights (see Fig. \ref{fig:setup} and Fig. \ref{fig:Th_spectrum}) with the aim to resolve the spatial dependency of the $A/E$ parameter.

\subsubsection{Correlation between $A/E$ and rise time}
\label{subsubsec:PSD_homogeneity}

\begin{figure}
	\includegraphics[scale=0.45]{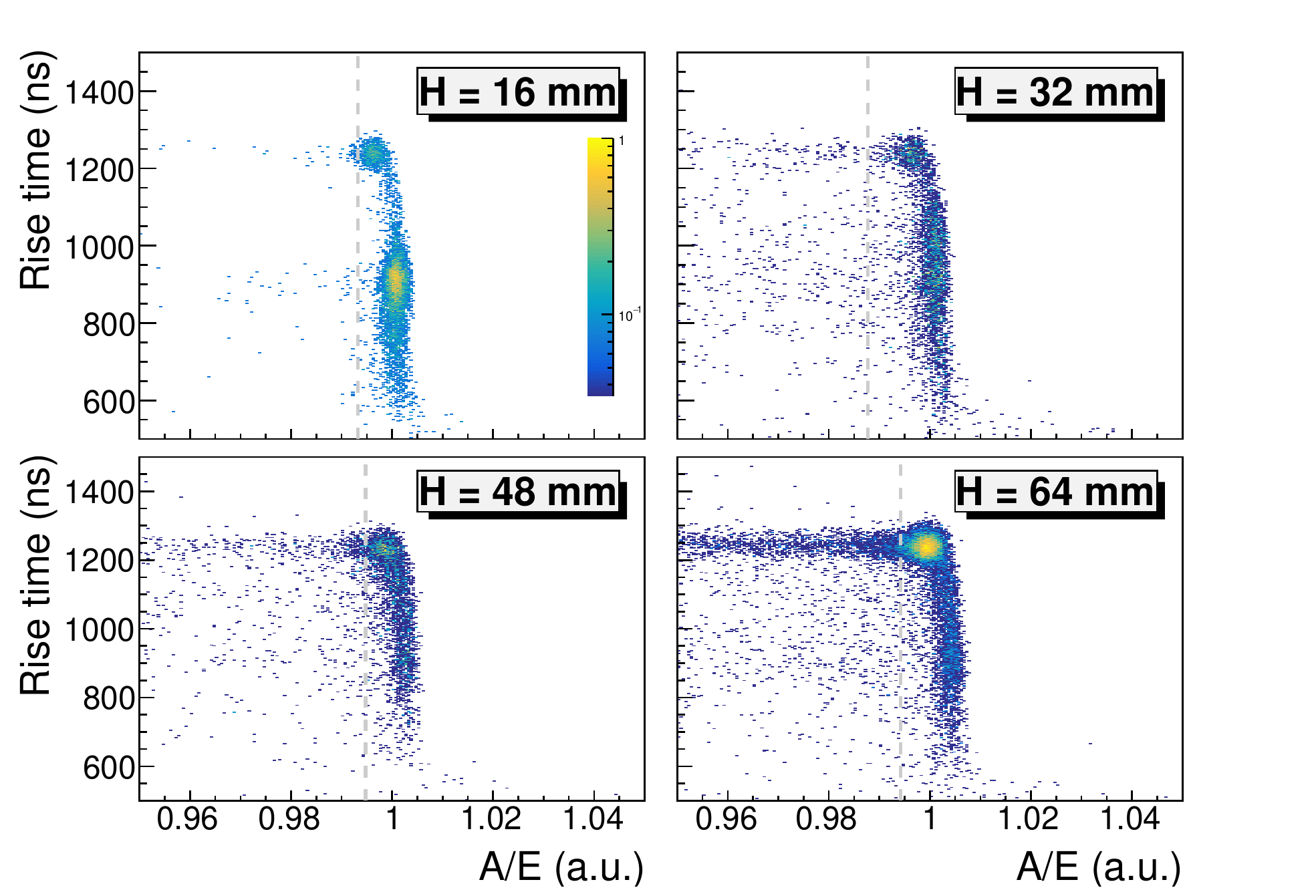}
	\caption{Correlation of $A/E$ with the signal rise time [0.5\%--90\%] of detector 50A for indicated heights $H$ of the lateral collimator.
	The p+ contact is at $H=0$~mm.
	The gray dashed lines show the $A/E$ cut position.
	These four datasets correspond to \Tl DEP events from the highly collimated \Th source.}
	\label{fig:AoE_scan}
\end{figure}

From Sect. \ref{subsec:SurfaceScan}, the $A/E$ ratio of SSEs is expected to be distributed around two values from energy depositions that happen close to and far away from the p+ contact.
These two values are close but could be resolved in the low electronic noise environment of the vacuum cryostat.
On Fig. \ref{fig:AoE_scan}, the lateral scan data show the two SSE $A/E$ populations from the \Tl DEP line of detector 50A: short rise time with high $A/E$ and long rise time with low $A/E$.
At the closest position to the p+ contact ($H = 16$ cm), the dominant SSE population has a rise time between 800~ns and 950~ns.
This is about 50~ns shorter than what is observed in the lateral \Am source scan (Fig. \ref{fig:AmScan}c).

This observation can be explained in part by the fact that the 2615~keV $\gamma$ rays penetrate deeper in the detector and hence probe a region on average closer to the p+ contact.
Additionally, \Am\ source lateral scans are performed along a single detector orientation as opposed to flood measurements that probes a large fraction of the detector bulk.
As a result, scans produce pulses with a specific charge velocity, sensitive to the closest crystal axis, that is not representative of the average bulk velocity.
At higher positions up to the top ($H = 64$~mm), the main population is moving towards longer rise times of about 1200~ns.
The shift between the two $A/E$ populations is found to be 0.4\% for this detector which is compatible with the observed variation in \Am source measurements.

\subsubsection{Background rejection}
\label{subsubsec:PSD_rej}

\begin{figure}
	\centering
	\includegraphics[width=\linewidth]{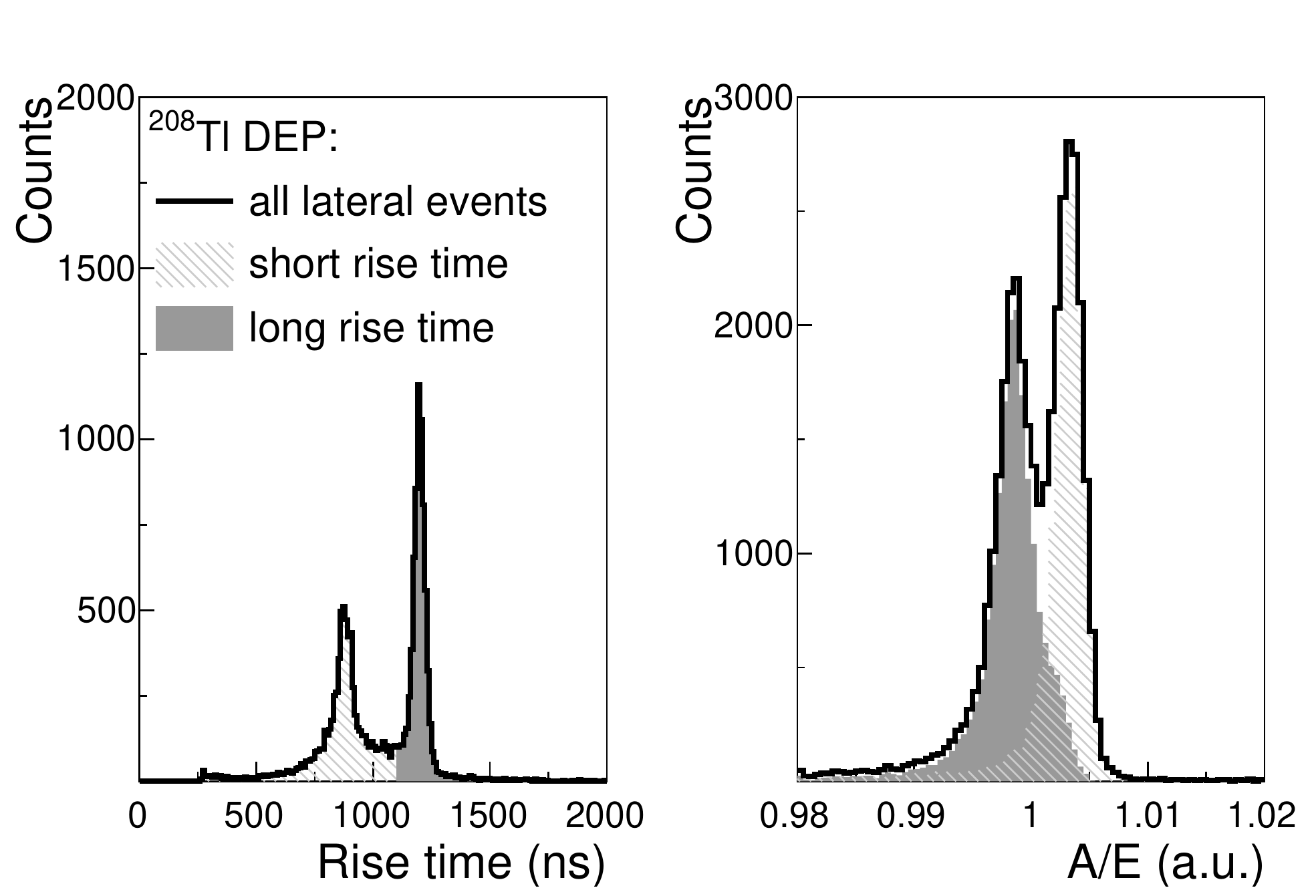}
	\caption{Left (right): Rise time ($A/E$) distribution of \Tl DEP events obtained from the lateral \Th source flood measurement with detector 50A.
		The \textit{lateral}, \textit{short} and \textit{long} analysis datasets selection are emphasized.}
	\label{fig:IC50A_RT_DEP}
\end{figure}

The rejection of MSEs for the two rise time populations (Fig. \ref{fig:IC50A_RT_DEP}) was studied by considering the lateral \Th source flood measurement (Fig. \ref{fig:setup}).
The effect of charge cloud diffusion is here most enhanced due to well distributed event statistics across the detector height.
The PSD cut is set on the low $A/E$ distribution side to accept 90\% of the \Tl DEP events.
Various peaks of interest featuring variable proportions of SSEs and MSEs have been investigated.
Their survival fractions after applying the cut, for four different analysis configurations, are summarized in Table \ref{tab:IC50A_survival_fraction} for detector 50A.
The \textit{lateral} analysis dataset combines all data without any correction or selection.
In addition, the \textit{corrected} analysis shows the PSD performance when the linear rise time dependence of $A/E$ (cf. Fig. \ref{fig:AoE_scan}) is corrected for.
This energy dependent linear correction aims to cancel the position dependence of the SSE $A/E$ by empirically minimizing the width of the SSE band $A/E$ distribution.
This is done by 1) fitting its rise time dependence at many Compton continuum energy regions above 1~MeV:
\begin{equation}
	A/E(RT) = a(E) + b(E) \times RT
\end{equation}
and, 2) applying this correction to all events:
\begin{equation}
	A/E^{\textrm{corr}}_{i} = (A/E_i - b(E) \times RT_i) / a(E)
\end{equation}
The \textit{short RT} and \textit{long RT} analysis datasets comprise only the events that fall into the short and long rise time population, respectively (cf. Fig. \ref{fig:IC50A_RT_DEP}).
Despite a double $A/E$ peak for SSEs, the \textit{lateral} analysis shows excellent rejection of MSEs (32\% acceptance of Compton continuum at \qbb) as compared to the \gerda\ BEGe detectors ($\sim$40\%)~\cite{bib:BEGe2019}.
Similar performance is observed if the linear rise time correction is applied to $A/E$.
However, this position dependence correction comes at the expense of a better understanding of its energy dependence and will require additional studies on other classes of events found in \gerda~\cite{bib:Bkg2020}.
The \textit{short RT} and \textit{long RT} analysis quantify the ability of our PSD method to discriminate SSEs and MSEs in the lower and upper parts of the detector, respectively.
As expected, it does perform significantly better for the upper part as the separation in time between the hits is on average larger.
Care must be taken not to overinterpret this qualitative observation as MSEs do feature, by definition, longer rise time.

\begin{table}[htbp]
	\centering
	\caption{
		Survival fractions (in \%) and their statistical uncertainty obtained with detector 50A and the low-sided PSD cut at indicated $\gamma$ line regions and the Compton continuum around \qbb\ (CC@\qbb).}
	\begin{tabular}{c|c|c|c|c} 
		Event class         & \textit{lateral} & \textit{corrected}   & \textit{short RT} & \textit{long RT} \\ 
		\hline
		\Tl\ DEP             & 90(2)  	 & 90(2)     & 90(3)   	& 90(3)   \\ 
		\Bi\ FEP  		     & 9.3(4)    & 8.3(4)    & 10.2(7) & 5.9(5)  \\	 
		CC @ \qbb 		  & 33.4(6)  & 33.8(6)   & 37.1(9)  & 30.6(8) \\	
		\Tl\ SEP             & 6.3(4) 	 & 6.2(4)    & 8.3(6)  	& 4.3(5)   \\ 			
		\Tl\ FEP  		     & 8.05(8)	& 8.47(8)   & 11.0(1)  & 6.8(1) \\ 
	\end{tabular}
	\label{tab:IC50A_survival_fraction}
\end{table}

The background rejection performance of all five IC detectors for the top \Th source measurements, without correction, is summarized in Table \ref{tab:top_survival_fraction}.
The values are comparable to the ones obtained with BEGe and PPC detectors~\cite{bib:BEGe2019,bib:MJDpsd}.

\begin{table}[h!]
	\centering
	\caption{Survival fractions (in \%) and their statistical uncertainty obtained with top \Th source position and the low-sided PSD cut at indicated energy regions.}
	\begin{tabular}{c|c|c|c|c|c} 
		
		Event class        &  48A    	& 48B      &  50A        &  50B      & 74A       \\ \hline
		\Tl DEP     	     &  90(2)     &  90(1)    &  90(1)      &  90(1)     & 90(3)     \\ 
		\Bi FEP  		     &  6.9(9)    &  8.0(5)	  &  8.6(4)     &  7.9(4)    & 12.1(9)  \\ 
		CC @ \qbb  		  &  37.1(8)   &  35.8(4) &  35.7(6)   & 36.6(5)   & 34.5(8)   \\ 
		\Tl SEP  		     &  6.4(7)    & 6.6(3)    &  6.5(3)     &  5.6(3)    & 7.6(6)    \\ 
		\Tl FEP  	     	 &  9.8(1)	  & 8.70(7)   &  9.28(8)   &  8.51(7)  &  -          \\ 
	\end{tabular}
	\label{tab:top_survival_fraction}
\end{table}

\section{Long-term performance in the GERDA cryostat}
\label{sec:Operation}

From July 2018 to November 2019, the five IC detectors have been operated within the \gerda\ LAr cryostat.
Among these detectors, detector 48B could not be fully depleted due to high leakage current over the entire data taking and has been biased with [3000 -- 3200]~V only.
As a consequence it was used in anti-coincidence mode only and excluded from the high level \bb\ decay analysis.
The averaged FWHM energy resolution at \qbb\ of the four others, observed during 18 months of operation, was 2.9(1)~keV.
During this data taking period, the background rate of $\alpha$ particles was studied to validate our detector handling procedure~\cite{bib:Bkg2020}, from detector production to integration in LAr.
We observed a constant rate of 4.5 $\alpha$ particles per month.
This is compatible with the one from BEGe detectors before the upgrade if normalized by the total p+ contact surface\footnote{This is the only location where $\alpha$'s can deposit their energy into the detector bulk.}.
It demonstrates the successful handling of the detectors during their fabrication and integration.
The weekly \Th source calibrations were used to study the PSD performance of these detectors in a similar manner as described in~\cite{bib:GERDA2013}.
The much higher electronics noise did not allow to resolve the two SSE populations shown in Fig. \ref{fig:IC50A_RT_DEP}.
While accepting 90\% of the \Tl DEP events with the low-side $A/E$ cut, the survival fractions of \Tl single escape peak (SEP) events was 7.4(7)\% and 35(2)\% for CC~@~\qbb\ on average. 
This is, respectively, 2\% and 7\% better than the achieved rejection performance with BEGe detectors~\cite{bib:Agostinieaav8613}.
None of the high energy $\alpha$'s survived the high $A/E$ cut, set to +3 standard deviations of the $A/E$ SSE band distribution.

\begin{figure*}
	\centering
	\includegraphics[width=\linewidth]{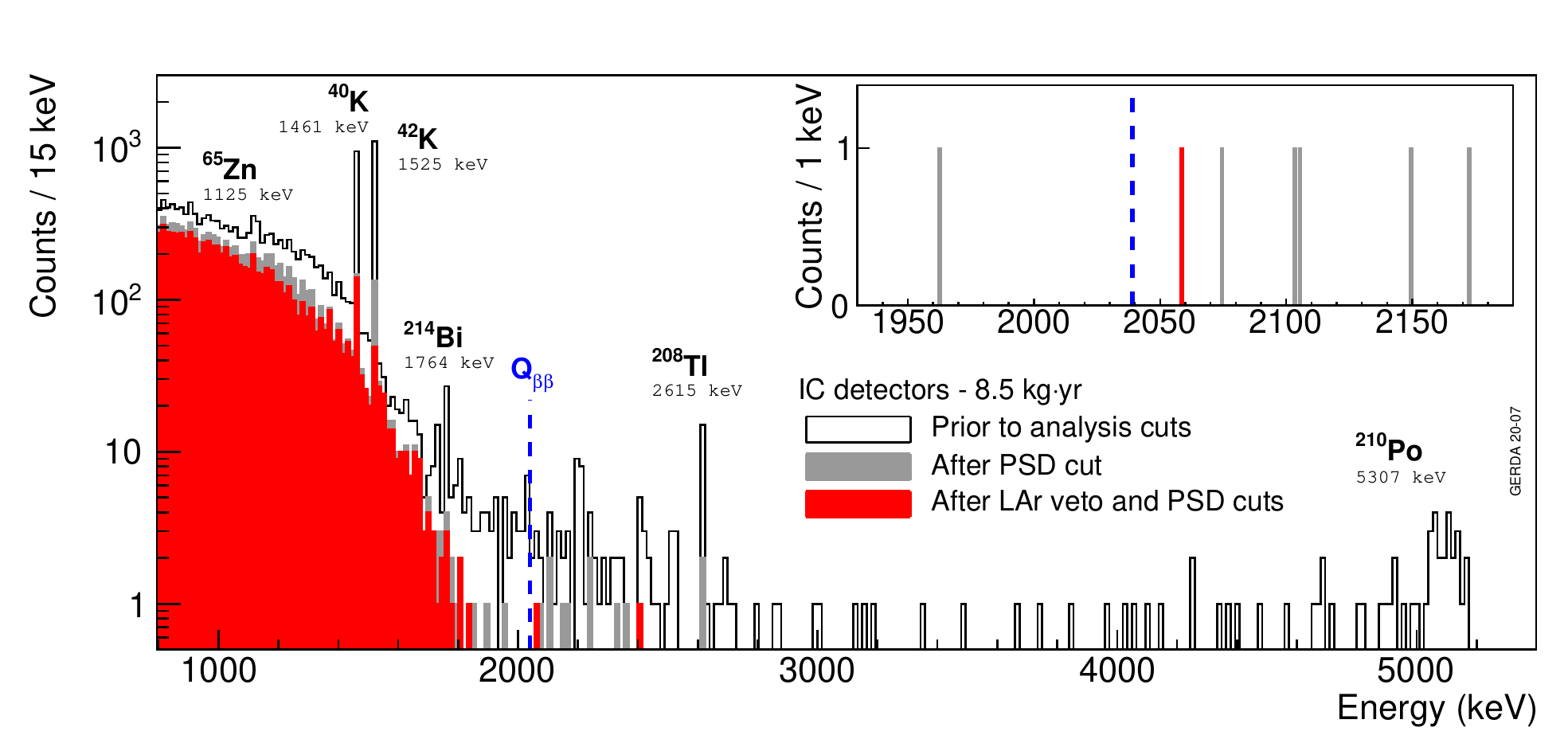}
	\caption{Spectrum measured with the IC detectors at the exposure of \expo\ in \gerda\ prior to and after indicated analysis cuts. 
		The inset shows a zoom in the background analysis window. 
		The only surviving event at 2058.9~keV was recorded with detector 74A on October 9, 2018, 01:09:14 (UTC).}
	\label{fig:phySpectrum}
\end{figure*}

Fig. \ref{fig:phySpectrum} shows the combined physics spectrum of the four working IC detectors at the cumulated exposure of \expo, before and after LAr veto and PSD cuts.
For a detailed description of the various features (2$\nu\beta\beta$ decay continuum, $\gamma$ lines from $^{40}$K and $^{42}$K and high energy $\alpha$'s) we refer to~\cite{bib:PRL2020,bib:Bkg2020}.
An additional $\gamma$ line was found on top of the usual energy spectrum at 1125~keV before analysis cuts.
This peak originates from $^{65}$Zn which produces a 1115.5~keV EC $\gamma$ accompanied by a 10~keV deexcitation X-ray.
This cascade is however highly suppressed by the PSD cut.
Such a detector bulk contamination has already been reported for some of the enriched \gerda\ coaxial detectors in~\cite{bib:BALYSH1994176} at the beginning of the Heidelberg-Moscow experiment.
With a half-life of 244 days, this $\gamma$ line had not been observed in previous \gerda\ datasets since coaxial and BEGe detectors had been stored at least 2 years underground before being deployed.

The final \gerda\ analysis led to a single event surviving the cuts in the [1930-2190]~keV region of interest (see inset of Fig. \ref{fig:phySpectrum}) which, with the omission of the two lines at 2104 $\pm$ 5~keV from \Tl SEP and 2119 $\pm$ 5~keV from $^{214}$Bi, was used to deduce the background index~\cite{bib:PRL2020}.
For the total exposure of 8.5~kg.yr the value is $\bidx \times 10^{-4} \ \textrm{counts}/(\textrm{keV} \cdot \textrm{kg} \cdot \textrm{yr})$.

\section{Conclusion}

Five enriched inverted coaxial \Ge detectors have been characterized in terms of surface response, energy resolution and background rejection efficiency.
High performance level was found for these last two critical items despite a peculiar distribution of the PSD parameter $A/E$ across the detector bulk.
After the \gerda\ upgrade in spring 2018, an exposure of \expo\ has been validated until fall 2019.
The standard analysis of \gerda\ reported a background rejection compatible with that of BEGe detectors for the $\alpha$ background and slightly improved for multi-site events at \qbb, while the active mass of the IC detectors is up to a factor 3 larger.
This is the first successful and stable long-term operation of IC detectors in LAr.
These very promising results serve as a proxy for estimating the PSD performance in ultra-low background environment for the upcoming \legend-200 experiment, currently under preparation at the INFN Laboratori Nazionali del Gran Sasso.

\begin{acknowledgements}
The transnational access scheme (EUFRAT) of the European Commission's Joint Research Centre in Geel is gratefully acknowledged.
The GERDA experiment is supported financially by
the German Federal Ministry for Education and Research (BMBF),
the German Research Foundation (DFG),
the Italian Istituto Nazionale di Fisica Nucleare (INFN),
the Max Planck Society (MPG),
the Polish National Science Centre (NCN),
the Foundation for Polish Science (TEAM/2016-2/17),
the Russian Foundation for Basic Research,
and the Swiss National Science Foundation (SNF).
This project has received funding/support from the European Union's
Horizon 2020 research and innovation programme under
the Marie Sklodowska-Curie grant agreements No 690575 and No 674896.
This work was supported by the Science and Technology Facilities Council (ST/T004169/1).
The institutions acknowledge also internal financial support.
The GERDA Collaboration thanks the directors and the staff of
the LNGS for their continuous strong support of the GERDA experiment.
We are also thankful to David Radford (ORNL) for helpful discussions.
\end{acknowledgements}


\end{document}